\documentclass[a4paper,11pt]{article}
\pdfoutput=1 % if your are submitting a pdflatex (i.e. if you have
\usepackage{jheppub}
\usepackage{cite} % for details on the use of the package, please
\newcommand{\cC}{\mathcal{C}}

\newcommand{\cT}{\mathcal{T}}
\newcommand{\cS}{\mathcal{S}}
\newcommand{\cM}{\mathcal{M}}
\newcommand{\cQ}{\mathcal{Q}}
\newcommand{\cPhi}{\varPhi}
\newcommand{\cH}{\mathcal{H}}
\newcommand{\cG}{\mathcal{G}}

\allowdisplaybreaks

\title{
Higher derivative corrections to black brane thermodynamics and the weak gravity conjecture}

\author{Toshifumi Noumi}

\author{and Hibiki Satake}

\affiliation{Department of Physics, Kobe University, Kobe 657-8501, Japan}

\emailAdd{tnoumi@phys.sci.kobe-u.ac.jp}

\emailAdd{hsatake@stu.kobe-u.ac.jp}

\preprint{KOBE-COSMO-22-15}

\abstract{
We study higher derivative corrections to black brane thermodynamics and their implications for the weak gravity conjecture for $p$-form gauge fields. In particular we show that higher derivative corrections decrease tension-to-charge ratios of extremal black branes as implied by the weak gravity conjecture, if four-derivative couplings follow scattering positivity bounds. We also demonstrate that entropy corrections in the micro canonical ensemble are positive under the same assumptions. This extends earlier works in the Einstein-Maxwell theory to $p$-form gauge fields in general spacetime dimensions.}

\begin{document}
\setcounter{tocdepth}{2}
\maketitle
\flushbottom

\section{Introduction}
\setcounter{equation}{0}

The weak gravity conjecture (WGC) claims that consistent quantum gravity theories should contain a charged state whose mass-to-charge ratio is smaller than unity in an appropriate unit~\cite{Arkani-Hamed:2006emk}. The conjecture if true has various implications for cosmology and particle physics, and so it has been studied intensively both theoretically and phenomenologically in the Swampland program~\cite{Vafa:2005ui} (see Refs.~\cite{Brennan:2017rbf,Palti:2019pca,vanBeest:2021lhn,Harlow:2022gzl} for review articles).

\medskip
An important question in this context is at which scale there appears the charged sate required by the WGC. As illustrated in the original papaer~\cite{Arkani-Hamed:2006emk}, string theory typically has charged states satisfying the WGC bound at various scales, both below and above the Planck scale. For example, for $1$-form gauge fields, the WGC bound is saturated by macroscopic extremal black holes in the Einstein gravity. As the black hole mass decreases, the spacetime curvature increases and so higher derivative corrections become relevant. As depicted in Fig.~\ref{corrections}, the extremal curve (for non-supersymmetric black holes) is typically pushed down monotonically by the corrections. This feature has been tested in string theory examples~\cite{Arkani-Hamed:2006emk,Natsuume:1994hd,Kats:2006xp,Cano:2019oma,Cano:2019ycn,Cano:2021nzo,Ma:2021opb} and studied in the effective field theory framework, e.g., from the viewpoints of scattering amplitudes and black hole thermodynamics~\cite{Cheung:2018cwt,Hamada:2018dde,Bellazzini:2019xts,Charles:2019qqt,Jones:2019nev,Loges:2019jzs,Goon:2019faz,Cremonini:2019wdk,Chen:2020rov,Loges:2020trf,Bobev:2021oku,Arkani-Hamed:2021ajd,Cremonini:2021upd,Aalsma:2021qga}.

\medskip
If we extrapolate the extremal curve down to the Planck scale, we expect a transition from black holes to microscopic objects. In string theory, excited strings would be responsible for this~\cite{Bowick:1985af,Susskind:1993ws,Horowitz:1996nw}. This perspective was illustrated in heteoric string examples~\cite{Arkani-Hamed:2006emk} and then further studied, e.g., using modular invariance~\cite{Heidenreich:2016aqi,Montero:2016tif,Lee:2018urn,Lee:2019tst,Aalsma:2019ryi,Klaewer:2020lfg}. There are also some attempts to constrain the spectrum of light particles well below the Planck scale from consistency of gravitational scattering amplitudes~\cite{Cheung:2014ega,Andriolo:2018lvp,Chen:2019qvr,Alberte:2020bdz,Aoki:2021ckh,Noumi:2021uuv,Noumi:2022zht}. The overall picture that there exist a tower of WGC states at various scales is summarized by the subLattic/Tower WGC~\cite{Heidenreich:2015nta,Heidenreich:2016aqi,Montero:2016tif,Andriolo:2018lvp}.
 
\begin{figure}[t]
\begin{center}
\includegraphics[width=80mm, bb=0 0 1024 768]{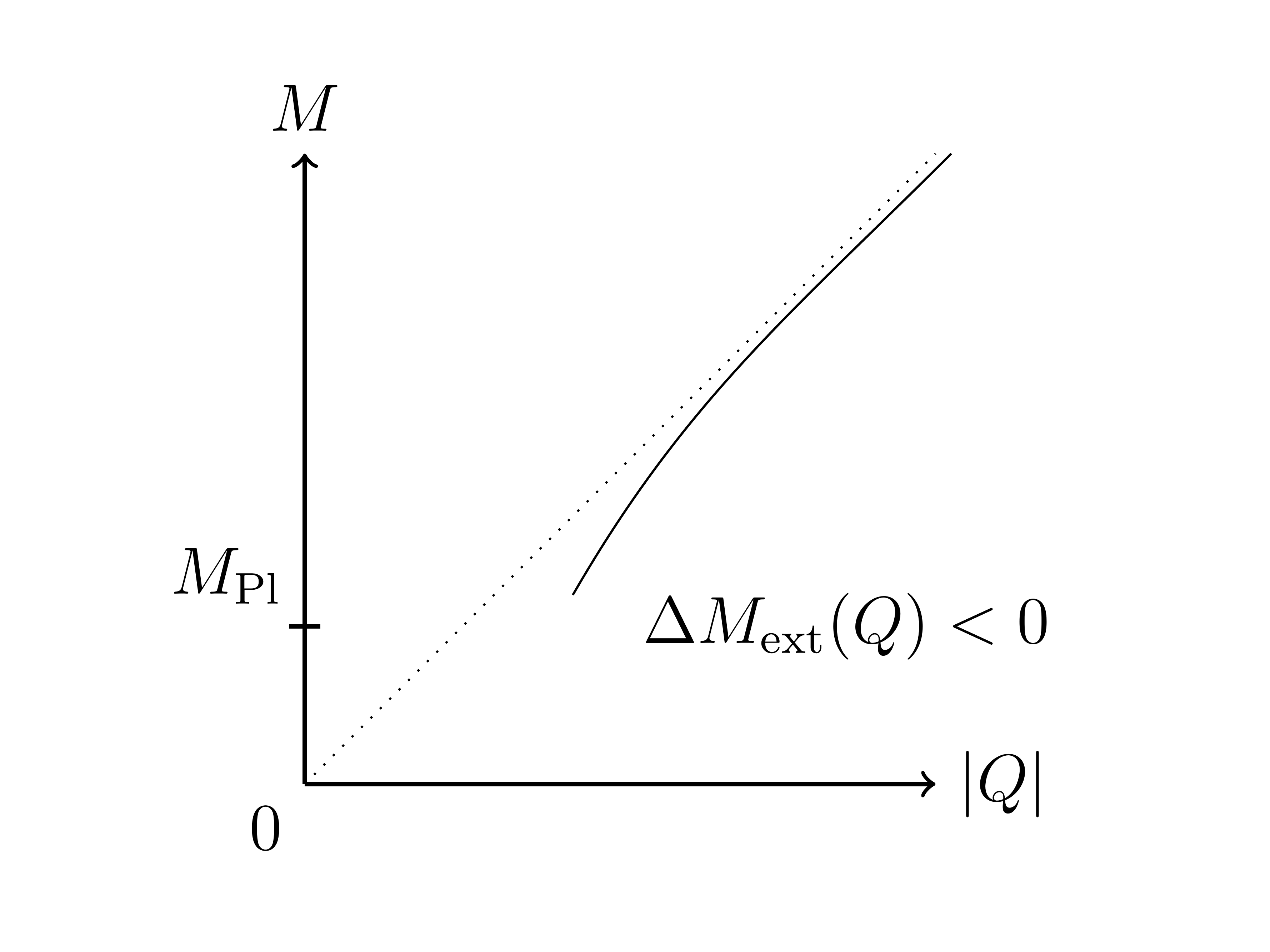}
\end{center}
\vspace{-5mm}
\label{corrections}
\caption{Monotonic behavior of extremal curve favored by the WGC: The dots stand for the extremal curve in the Einstein gravity. If higher derivative corrections decrease the mass/tension of extremal black holes/branes $\Delta M_{\rm ext}(Q)<0$, large black holes/branes satisfy the WGC bounds.}
\label{corrections}
\end{figure}

\medskip
In this paper we would like to generalize the story to $p$-form gauge fields. More specifically, we study higher derivative corrections to extremal conditions of charged black branes from a thermodynamic perspective and discuss under which conditions the extremal curve is pushed down as implied by the WGC (see also Ref.~\cite{Andriolo:2020lul} for the axionic WGC). We show that the WGC is satisfied if four-derivative couplings follow scattering positivity bounds~\cite{Adams:2006sv} (with a caveat on positivity in the presence of gravity mentioned later). We also demonstrate that entropy corrections in the micro canonical ensemble are positive under the same assumptions. This extends earlier works in the Einstein-Maxwell theory to $p$-form gauge fields in general spacetime dimensions.

\medskip
The paper is organized as follows. In Sec.~\ref{Sec:thermo} we review black brane thermodynamics in the Einstein gravity. In Sec.~\ref{Sec:corrections} we study higher derivative corrections to the thermodynamics and their implications for the WGC. Discussion of our results is given in the final section. Technical details are collected in appendices.

\section{Black brane thermodynamics}
\label{Sec:thermo}

We begin by a brief review of black brane solutions and their thermodynamics in the Einstein gravity to summarize our notation. 

\subsection{Black brane solution in Einstein gravity}

We consider a $(p+1)$-form gauge field $A$ coupled to gravity in $d$-dimensional spacetime:
\begin{equation}
	\mathcal{I}_0 = \frac{1}{2\kappa^{2}} \int d^{d}x \sqrt{-g}R - \frac{1}{2e^{2}}\int d^{d}x \sqrt{-g}F^{2}
+\mathcal{I}_{\rm GHY}\,,	\label{action1}
\end{equation}
where $F=dA$ is the $(p+2)$-form field strength and $F^2\equiv\frac{1}{(p+2)!}F_{\mu_1\ldots\mu_{p+2}}F^{\mu_1\ldots\mu_{p+2}}$. $\mathcal{I}_{\rm GHY}$ represents the Gibbons-Hawking-York term, whose explicit form will be given later when necessary. This system accommodates black $p$-brane solutions of the form~\cite{Horowitz:1991cd},
\begin{align}
ds^{2} &=f_{-}(r)^{\frac{2}{p+1}}\left[
-\frac{f_{+}(r)}{f_-(r)}dt^2+\delta_{ij}dy^{i}dy^{j}
\right]
+ \frac{dr^{2}}{f_{+}(r)f_{-}(r)}+r^{2}d\Omega_{q}^{2}
\,,\label{BB1}	\\
F &= \frac{e}{\kappa}\frac{\sqrt{\frac{(p+q)(q-1)}{p+1}}(r_{+}r_{-})^{\frac{q-1}{2}}}{r^{q}}dt \wedge dy^{1} \wedge \cdots \wedge dy^{p} \wedge dr\,,\label{BB2}
\end{align}
where $q\equiv d-p-2$ and $f_{\pm}(r) \equiv 1- \left(\frac{r_{\pm}}{r}\right)^{q-1}$. We assume $q\geq2$ throughout the paper. Also, $t$ and $y^i$ ($i=1,\ldots,p$) are coordinates along the $p$-brane and $r$ is the radial coordinate.  $\Omega_q$ collectively represents coordinates of a $q$-dimensional unit sphere normal to the electric flux. An explicit form of $d\Omega_q^2$ is
\begin{align}
d\Omega_q^2=d\theta_1^2+\sin^2\theta_1d\theta_2^2+\sin^2\theta_1\sin^2\theta_2d\theta_3^2+\cdots+\left(\prod_{i=1}^{q-1}\sin^2\theta_i\right)d\theta_q^2\,.
\end{align}
The solution is parametrized by the location of the two horizons $r=r_\pm$.

\paragraph{Extremal condition.}

To identify the electric charge of the $p$-brane, it is convenient to write down the Hodge dual of the field strength:
\begin{align}
\star F=\frac{e}{\kappa}\sqrt{\tfrac{(p+q)(q-1)}{p+1}}(r_{+}r_{-})^{\frac{q-1}{2}}\omega\,,
\end{align}
where $\omega$ is the volume form of a $q$-dimensional unit sphere given by
\begin{align}
\omega=\left(\prod_{i=1}^{q-1}\sin^{i}\theta_{q-i} \right) d\theta_{1}\wedge\cdots\wedge d\theta_{q}\,.
\end{align}
Then, the electric charge (subject to the Dirac quantization $Q\in \mathbb{Z}$) reads
\begin{align}
	Q &= \frac{1}{e^{2}} \int_{S^q} \star F=
	\frac{V_q}{e\kappa}\sqrt{\tfrac{(p+q)(q-1)}{p+1}}(r_{+}r_{-})^{\frac{q-1}{2}}\,,
	\label{charge}
\end{align}
where $V_q$ is the volume of a $q$-dimensional unit sphere, i.e.,
\begin{align}
V_q=\int_{S^q}\omega=\frac{2\pi^{\frac{q+1}{2}}}{\Gamma(\frac{q+1}{2})}\,.
\end{align}
On the other hand, the ADM tension reads~\cite{Lu:1993vt}
\begin{align}
M &= \frac{V_q}{\kappa^2}
\frac{p+q}{p+1}
\left[r_{-}^{q-1}+\frac{q(p+1)}{2(p+q)}\left(r_{+}^{q-1}-r_{-}^{q-1}\right)\right]\,.	\label{tension}
\end{align}
Now, the extremal bound follows from $r_+\geq r_-$ as
\begin{align}
\left|\frac{M}{Q}\right|\geq\sqrt{\frac{p+q}{(p+1)(q-1)}}\frac{e}{\kappa}\,.
\label{ExtremalCondition}
\end{align}
The WGC requires a charged state whose tension-to-charge ratio is smaller than that of the macroscopic extremal black brane~\cite{Arkani-Hamed:2006emk}, i.e.,
\begin{align}
\left|\frac{M}{Q}\right|\leq\sqrt{\frac{p+q}{(p+1)(q-1)}}\frac{e}{\kappa}\,.
\end{align}
In the next section we study higher derivative corrections to the extremal condition~\eqref{ExtremalCondition} and discuss under which conditions black branes can play a role of the WGC state.

\subsection{Black brane thermodynamics}

Next we discuss the first law of the black brane solution (\ref{BB1})-(\ref{BB2}). For this, it is convenient to introduce the Euclidean time $t_E=it$ and the corresponding Euclidean solution:
\begin{align}
ds^{2} &=f_{-}(r)^{\frac{2}{p+1}}\left[
\frac{f_{+}(r)}{f_-(r)}dt_E^2+\delta_{ij}dy^{i}dy^{j}
\right]
+ \frac{dr^{2}}{f_{+}(r)f_{-}(r)}+r^{2}d\Omega_{q}^{2}
\,,\label{BBE1}	\\
F &= -i\frac{e}{\kappa}\frac{\sqrt{\frac{(p+q)(q-1)}{p+1}}(r_{+}r_{-})^{\frac{q-1}{2}}}{r^{q}}dt_E \wedge dy^{1} \wedge \cdots \wedge dy^{p} \wedge dr\,.\label{BBE2}
\end{align}
Then, the Hawking temperature is determined by requiring that the metric~\eqref{BBE1} has no conical singularity at the horizon $r=r_+$ as
\begin{equation}
	T = \frac{q-1}{4\pi}\frac{1}{r_{+}}\left[1-\left(\frac{r_{-}}{r_{+}}\right)^{q-1}\right]^{\frac{1}{p+1}}\,.	\label{temperature}
\end{equation}
Similarly, the Bekenstein-Hawking entropy per unit brane volume\footnote{The brane volume $\mathcal{V}=\int d^py$ is evaluated at the infinity $r=\infty$.} is given by the horizon area as
\begin{equation}
	S = \frac{2\pi V_q}{\kappa^{2}}r_{+}^{q}\left[1-\left(\frac{r_{-}}{r_{+}}\right)^{q-1}\right]^{\frac{p}{p+1}}\,.	\label{entropy}
\end{equation}
The electric potential is the value of the gauge potential at the event horizon,
\begin{equation}
	\Phi = \frac{e}{\kappa}\sqrt{\frac{p+q}{(p+1)(q-1)}}\left(\frac{r_{-}}{r_{+}}\right)^{\frac{q-1}{2}}	\,.	\label{potential}
\end{equation}
Note that they satisfy the Smarr-like formula $M=\frac{q}{q-1}TS+Q\Phi$. In the microcanonical ensemble, where the tension $M$ and the charge $Q$ are identified as independent variables, we confirm the following first law of black brane thermodynamics~\cite{Townsend:2001rg}:
\begin{equation}
	dM = TdS + \Phi dQ\,.	\label{1stlaw}
\end{equation}

\subsection{Other ensembles}

In the WGC context, we are interested in the tension-to-charge ratio of black branes of zero temperature, so that it is convenient to work in the canonical ensemble. Also, the grand canonical ensemble naturally appears in the Euclidean path integral analysis employed in the next section to study higher derivative corrections. Having this in mind, we introduce several thermodynamic ensembles and the corresponding potentials.

\paragraph{Generality.}

The Helmholtz free energy $H=H(T, Q)$ in the canonical ensemble and the Gibbs free energy $G=G(T, \Phi)$ in the grand canonical ensemble are defined by
\begin{equation}
	H = M - TS\,, \qquad G = H - Q\Phi\,.
\end{equation}
The first law~(\ref{1stlaw}) in the microcanonical ensemble then implies
\begin{equation}
	dH = -S dT + \Phi dQ\,,\qquad dG = -SdT - Qd\Phi\,.	\label{1}
\end{equation}
We emphasize that these relations hold in general (beyond the Einstein gravity), as long as the first law~(\ref{1stlaw}) in the micro canonical ensemble is satisfied. Indeed, the Wald entropy~\cite{Wald:1993nt} is defined such that the first law is satisfied, which provides a natural generalization of the Bekenstein-Hawking entropy formula in higher derivative theories.

\paragraph{Rescaled variables.}

For notational simplicity, it is convenient to introduce rescaled thermodynamic quantities as follows:
\begin{align}
&\cM=\frac{p+1}{p+q}\frac{\kappa^2}{V_q}M	\,,
\quad
\cT=\frac{4\pi}{q-1}T
\,,
\quad
\cS=\frac{(p+1)(q-1)}{4\pi(p+q)}\frac{\kappa^2}{V_q}S\,,
\label{rescaling1}
\\
&\cQ=\sqrt{\frac{p+1}{(p+q)(q-1)}}\,\frac{e\kappa}{V_q}Q\,,
\quad
\cPhi=\sqrt{\frac{(p+1)(q-1)}{p+q}}\,\frac{\kappa}{e}\Phi\,.	\label{rescaling2}
\end{align}
Correspondingly, we define the rescaled free energies as
\begin{align}
\cH=\cM-\cT \cS=\frac{p+1}{p+q}\frac{\kappa^2}{V_q}H\,,
\quad
\cG=\cH-\cQ\cPhi=\frac{p+1}{p+q}\frac{\kappa^2}{V_q}G\,.
\label{rescaling3}
\end{align}
Note that the first laws are in the standard form even after the rescaling:
\begin{align}
\label{1st_general}
d\cM=\cT d\cS+\cPhi d\cQ\,,
\quad
d\cH=-\cS d\cT+\cPhi d\cQ\,,
\quad
d\cG=-\cS d\cT-\cQ d\cPhi\,.
\end{align}

\paragraph{Grand canonical ensemble in Einstein gravity.}

For later use, we provide an explicit form of thermodynamic quantities in the Einstein gravity for the grand canonical ensemble, where the temperature $T$ and the electric potential $\Phi$ are independent thermodynamic variables. First, the electric charge $\cQ$, the ADM tension $\cM$, and the entropy density $\cS$ in the rescaled language are given by
\begin{align}
\label{cQ0}
\cQ_0(\cT,\cPhi)&=\cPhi\frac{(1-\cPhi^2)^{\frac{q-1}{p+1}}}{\cT^{q-1}}\,,
\\
\cM_0(\cT,\cPhi)
&=
\left[
\cPhi^2
+\frac{q(p+1)}{2(p+q)}(1-\cPhi^2)\right]
\frac{(1-\cPhi^2)^{\frac{q-1}{p+1}}}{\cT^{q-1}}\,,
\\
\cS_0(\cT,\cPhi)&=\frac{(q-1)(p+1)}{2(p+q)}
\frac{(1-\cPhi^2)^{\frac{p+q}{p+1}}}{\cT^{q}}\,.
\end{align}
Here and in what follows we use the subscript $0$ to indicate that the expression is for the Einstein gravity without higher derivative corrections. Accordingly, the rescaled Gibbs free energy reads
\begin{align}
\label{G0_standard}
\cG_0(\cT,\cPhi)=\cM_0(\cT,\cPhi)-\cT\cS_0(\cT,\cPhi)-\cPhi\cQ_0(\cT,\cPhi)=\frac{p+1}{2(p+q)}\frac{(1-\cPhi^2)^{\frac{p+q}{p+1}}}{\cT^{q-1}}\,.
\end{align}
As a consistency check, one may explicitly show that the first law~\eqref{1st_general} is satisfied, i.e.,
\begin{align}
\cQ_0(\cT,\cPhi)=-\frac{\partial \cG_0(\cT,\cPhi)}{\partial \cPhi}\,,
\quad
\cS_0(\cT,\cPhi)=-\frac{\partial \cG_0(\cT,\cPhi)}{\partial \cT}\,.
\end{align}

\section{Higher derivative corrections}
\label{Sec:corrections}

In this section we study higher derivative corrections to the black brane thermodynamics and their implications for the WGC. Assuming parity invariance, we consider the following Lorentzian effective action with the leading order corrections relevant to our problem\footnote{
When $p+1=4n$ ($n=1,2,\ldots$), the effective action contains three-derivative operators schematically of the form, $F^3$, but we ignore these terms because they do not modify the profile of electric brane solutions and their thermodynamics.
}:
\begin{align}
\label{Lorentzian_action}
\mathcal{I} &= \mathcal{I}_0+\Delta \mathcal{I}
\,,
\\
\label{Lorentzian_4}
\Delta \mathcal{I}&=\frac{\kappa^{2}}{e^{4}}\sum_{j}\alpha_{j}\int d^{d}x \sqrt{g}\left(F^{4}\right)_{j} + \frac{\beta}{e^{2}} \int d^{d}x \sqrt{g} FFW + \frac{\gamma}{\kappa^{2}} \int d^{d}x \sqrt{g}W^{2}\,,
\end{align}
where $\mathcal{I}_0$ is the Einstein gravity action~\eqref{action1} and $W_{\mu\nu\rho\sigma}$ is the Weyl tensor.
Other four-derivative operators, e.g., schematically of the form $(\nabla F)^2$, can be removed by field redefinition, hence this provides the most general (parity even) four-derivative action.
The first term in Eq.~\eqref{Lorentzian_4} denotes operators with four field-strengths and $j$ is a label for tensor structures. For example, for $p=0$, we have two independent tensor structures ($j=1,2$):
\begin{align}
\left(F_{\mu\nu}F^{\mu\nu}\right)^2\,,
\quad
F_{\mu\nu}F^{\nu\rho}F_{\rho\sigma}F^{\sigma\mu}\,.
\end{align}
Similarly, for $p=1$, we have three independent structures ($j=1,2,3$):
\begin{align}
(F_{\mu\nu\rho}F^{\mu\nu\rho})^2
\,,
\quad
F^{\mu\alpha\beta}F_{\mu}{}^{\gamma\delta}F^{\nu}{}_{\alpha\beta}
F_{\nu\gamma\delta}\,,
\quad
F^{\mu\alpha\beta}F_{\mu}{}^{\gamma\delta}F^{\nu}{}_{\alpha\gamma}
F_{\nu\beta\delta}\,.
\end{align}
The second term $FFW$ denotes an operator with two field-strengths and one Weyl tensor, for which there exists only one independent tensor structure:
\begin{align}
FFW=F_{\mu_1\mu_2\nu_1\nu_2\ldots \nu_p}F_{\mu_3\mu_4}{}^{\nu_1\mu_2\ldots \nu_p}W^{\mu_1\mu_2\mu_3\mu_4}\,.
\end{align}
The third term $W^2$ is $W^2=W_{\mu\nu\rho\sigma}W^{\mu\nu\rho\sigma}$.
Besides, higher derivative corrections to the boundary action are irrelevant in our analysis as long as $q\geq2$, essentially because higher derivative operators decay quickly far away from the brane. In the following, we study the effects of the four-derivative operators in Eq.~\eqref{Lorentzian_4} on the black brane thermodynamics.

\subsection{Outline}

\paragraph{Gibbs free energy from Euclidean path integral.}

In the thermodynamic analysis, we employ the Euclidean path integral formalism based on the following Euclidean action for the setup~\eqref{Lorentzian_action}:
\begin{align}
	I[\phi] &= I_0[\phi]+\Delta I[\phi]\,,
	\label{action2}
	\\
	I_0[\phi]&=-\frac{1}{2\kappa^{2}} \int_{\mathcal{M}} d^{d}x \sqrt{g}R+ \frac{1}{2e^{2}}\int_{\mathcal{M}} d^{d}x \sqrt{g}F^{2} - I_{\text{GHY}} \,,
	\\
	\Delta I &= -\frac{\kappa^{2}}{e^{4}}\sum_{j}\alpha_{j}\int d^{d}x \sqrt{g}\left(F^{4}\right)_{j} - \frac{\beta}{e^{2}} \int d^{d}x \sqrt{g} FFW - \frac{\gamma}{\kappa^{2}} \int d^{d}x \sqrt{g}W^{2}\,,
\end{align}
where $\phi=g_{\mu\nu},A$ collectively denotes the metric $g_{\mu\nu}$ and the $(p+1)$-form gauge field $A$. $I_{\text{GHY}}$ is the Gibbons-Hawking-York term \cite{York:1972sj, Gibbons:1976ue},
\begin{equation}
	I_{\text{GHY}} = \frac{1}{\kappa^{2}} \int_{\partial\mathcal{M}} d^{d-1}x \sqrt{h} \left(K-K_{0}\right)\,,
\end{equation}
where $\partial \mathcal{M}$ is the boundary of the spacetime manifold $\mathcal{M}$, $h$ is the induced metric on $\partial\mathcal{M}$, $K$ is the trace of the extrinsic curvature of $\partial\mathcal{M}$, and $K_{0}$ is that for flat space necessary to make the on-shell action finite.

\medskip
For the action~\eqref{action2}, it is appropriate to impose Dirichlet boundary conditions on the electric potential. Also, we impose periodic boundary conditions along the Euclidean time specified by the temperature $T$. Then, the Euclidean path integral gives the grand canonical ensemble, where the temperature $T$ and the electric potential $\Phi$ are thermodynamic variables. At the semiclassical level, the Gibbs free energy per unit brane volume reads
\begin{align}
\label{Gibbs_path_integral}
G=\frac{TI[\bar{\phi}]}{\mathcal{V}}\,,
\end{align}
where $\mathcal{V}=\int d^py$ is the brane volume evaluated at the infinity $r=\infty$ and $\bar{\phi}$ stands for the classical solution in the setup~\eqref{action2}. Other thermodynamic quantities follow from the Gibbs free energy based on the standard thermodynamic argument, so that our first task is to evaluate higher derivative corrections to the on-shell Euclidean action.

\paragraph{Higher derivative analysis.}

For this purpose, let us expand the solution $\bar{\phi}$ in the full theory
\eqref{action2} as
\begin{align}
\bar{\phi}=\bar{\phi}_0+\Delta \bar{\phi}\,,
\end{align}
where $\bar{\phi}_0$ is the solution in the Einstein gravity and $\Delta \bar{\phi}$ is the  higher derivative correction. Note that we are working in the grand canonical ensemble and so the solution $\bar{\phi}$ should be regarded as a function of the temperature $T$, the electric potential $\Phi$, and the Wilson coefficients $\alpha_i$, $\beta$, and $\gamma$. More explicitly,
\begin{align}
\Delta \bar{\phi}(T,\Phi):=\bar{\phi}(T,\Phi)-\bar{\phi}_0(T,\Phi)\,.
\end{align}
Accordingly, we can expand the on-shell action as
\begin{align}
I[\bar{\phi}]&=I_0[\bar{\phi}_0+\Delta \bar{\phi}]
+\Delta I[\bar{\phi}_0+\Delta\bar{\phi}]
\nonumber
\\
&\simeq I_0[\bar{\phi}_0]
+\int d^dx\left.\frac{\delta I_0}{\delta \phi}\right|_{\phi=\bar{\phi}_0}\Delta\bar{\phi}
+\Delta I[\bar{\phi}_0]\,,
\end{align}
where we kept the leading order correction only. A simple, but useful observation is that the equations of motion guarantee that the second term vanishes up to a boundary term. Also, the boundary term vanishes as long as $q\geq2$, essentially because the spacetime curvature and the electric flux decay quickly away from the brane. As a result, we find
\begin{align}
I[\bar{\phi}]&\simeq I_0[\bar{\phi}_0]
+\Delta I[\bar{\phi}_0]
\end{align}
at the leading order. In particular, we do not need to evaluate the correction $\Delta\bar{\phi}$ to the solution explicitly, which simplifies the analysis significantly~\cite{Reall:2019sah}. Then, the Gibbs free energy reads
\begin{align}
G(T,\Phi)=G_0(T,\Phi)+\Delta G(T,\Phi)
\end{align}
with
\begin{align}
G_0(T,\Phi)=\frac{T I_0[\bar{\phi}_0]}{\mathcal{V}}\,,
\quad
\Delta G(T,\Phi)=\frac{T \Delta I[\bar{\phi}_0]}{\mathcal{V}}\,.
\end{align}
Correspondingly, we define the rescaled correction $\Delta\cG$ as (see also Eq.~\eqref{rescaling3})
\begin{align}
\cG(\cT,\cPhi)=\cG_0(\cT,\cPhi)+\Delta 
\cG(\cT,\cPhi)
\,,
\quad
\Delta\cG(\cT,\cPhi)=\frac{p+1}{p+q}\frac{\kappa^2}{V_q}\Delta G(T,\Phi)\,.
\end{align}
Note that one can explicitly show that $\cG_0$ evaluated in this manner agrees with Eq.~\eqref{G0_standard}. Below, we first evaluate $\Delta \cG$ and then perform thermodynamic arguments.

\subsection{Gibbs free energy}

Now we compute the higher derivative corrections to the Gibbs free energy.

\paragraph{$F^4$ term.}

First, it is convenient to notice that for the Einstein gravity solution~\eqref{BBE1}-\eqref{BBE2} the $\alpha_j$ operators schematically of the form $F^4$ are simply
\begin{align}
\label{F^4}
\left(F^4\right)_j
=A_j \,(g^{t_Et_E}g^{rr}g^{y_1y_1}\cdots g^{y_py_p})^2(F_{t_Ery^{1}\ldots y^{p}})^4
=A_j \,(F_{try^{1}\ldots y^{p}})^4
\end{align}
with a combinatorial factor $A_j$. For example, for $p=0$, there are two operators,
\begin{align}
\left(F^4\right)_1=(F_{\mu\nu}F^{\mu\nu})^2
\,,
\quad
\left(F^4\right)_2=F_{\mu\nu}F^{\nu\rho}F_{\rho\sigma}F^{\sigma\mu}
\,,
\end{align}
and the corresponding combinatorial factors read
\begin{align}
A_1=4
\,,
\quad
A_2=2
\,.
\end{align}
Similarly, for $p=1$, we have
\begin{align}
\left(F^4\right)_1=(F_{\mu\nu\rho}F^{\mu\nu\rho})^2
\,,
\,\,
\left(F^4\right)_2=F^{\mu\alpha\beta}F_{\mu}{}^{\gamma\delta}F^{\nu}{}_{\alpha\beta}
F_{\nu\gamma\delta}\,,
\,\,
\left(F^4\right)_3=F^{\mu\alpha\beta}F_{\mu}{}^{\gamma\delta}F^{\nu}{}_{\alpha\gamma}
F_{\nu\beta\delta}\,,
\end{align}
and
\begin{align}
A_1=36
\,,
\quad
A_2=12
\,,
\quad
A_3=6
\,.
\end{align}
It is straightforward to compute the correction to the Gibbs free energy as
\begin{align}
\Delta G|_{F^4}&=-\Big(\sum_iA_i\alpha_i\Big) \frac{(p+q)^2(q-1)^2}{(p+1)^2}\frac{V_q}{\kappa^2}\int_{r_+}^\infty dr \frac{(r_+r_-)^{2(q-1)}}{r^{3q}}
\nonumber
\\*
&=-\Big(\sum_iA_i\alpha_i\Big) \frac{(p+q)^2(q-1)^2}{(p+1)^2(3q-1)}\frac{V_q}{\kappa^2}r_+^{q-3}\left(\frac{r_-}{r_+}\right)^{2(q-1)}
\nonumber
\\*
&=-\Big(\sum_iA_i\alpha_i\Big) \frac{(p+q)^2(q-1)^2}{(p+1)^2(3q-1)}\frac{V_q}{\kappa^2}\cT^{3-q}\cPhi^4(1-\cPhi^2)^{\frac{q-3}{p+1}}\,.
\end{align}
Here and in what follows, we use $O|_{F^4}$ to denote the contribution from the $\alpha_j$ couplings to the quantity $O$, and similarly for other higher derivative operators. In terms of the rescaled Gibbs free energy, we have
\begin{align}
\Delta \cG|_{F^4}=-\Big(\sum_iA_i\alpha_i\Big) \frac{(p+q)(q-1)^2}{(p+1)(3q-1)}\cT^{3-q}\cPhi^4(1-\cPhi^2)^{\frac{q-3}{p+1}}\,.
\end{align}

\paragraph{$FFW$ term.}

Other contributions can also be evaluated in a straightforward manner. First, the correction from the $FFW$ term to the rescaled Gibbs free energy reads\footnote{
In Appendix~\ref{App:curvature} for, we collect useful formulae including an explicit form of curvature tensors for the black brane solution~\eqref{BB1} in the Einstein gravity.}
\begin{align}
\label{delatGFFW}
\Delta \mathcal{G}_{FFW} &=
\beta (q-1)^{2}p!\left[\frac{2\mathcal{R}_{FFW}}{(p+1)(p+q+1)(3q-1)}
+(\cPhi^{-2}-1)
\right]
\mathcal{T}^{3-q}\varPhi^{4}\left( 1 - \varPhi^{2}\right)^{\frac{q-3}{p+1}}\,.
\end{align}
Here $\mathcal{R}_{FFW}$ is a $(p,q)$-dependent factor defined by
\begin{align}
\mathcal{R}_{FFW} &=p^2(2q-1)+p(5q-2)-(q^2-4q+1)\,.
\end{align}
Note that the second term in the brackets of Eq.~\eqref{delatGFFW} vanishes in the limit $\cPhi^2\to1$, which corresponds to the extremal limit in the Einstein gravity. Essentially because of this, it does not affect the extremal condition and the entropy of the $\cM=\cQ$ black branes as we explicitly see later.

\paragraph{$W^2$ term.}

Similarly, the correction from the $W^2$ term reads
\begin{align}
\label{delatWW}
\Delta \mathcal{G}_{W^2} &=
-\gamma\left[\frac{2(q-1)\mathcal{R}_{W^2}}{(p+1)^{2}(p+q+1)(3q-1)}
+\mathcal{O}(1-\cPhi^2)
\right]
\mathcal{T}^{3-q}\left( 1 - \varPhi^{2}\right)^{\frac{q-3}{p+1}}\,.
\end{align}
Here the second term in the brackets denotes subleading terms in the limit $\cPhi^2\to1$, whose explicit form is given in Appendix~\ref{app:GWW}. Similarly to the second term in the brackets of Eq.~\eqref{delatGFFW}, they do not affect the extremal condition and the entropy of the $\cM=\cQ$ black branes. Also, $\mathcal{R}_{W^{2}}$ is a $(p,q)$-dependent factor given by
\begin{align}
\mathcal{R}_{W^{2}} &= p^{3}q(2q-1)	+ 2p^{2}(q^3+2q-1)	+ p(3q^3-6q^2+12q-4) + 2(q^3-3q^2+4q-1) \,.
\end{align}

\paragraph{Summary of the subsection.}

Collecting all the above results and picking up the leading order term in the limit $\cPhi^2\to1$, we conclude that
\begin{align}
\label{DeltaG}
\Delta \mathcal{G} &=
-\Big[\cC
+\mathcal{O}(1-\cPhi^2)
\Big]
\mathcal{T}^{3-q}\left( 1 - \varPhi^{2}\right)^{\frac{q-3}{p+1}}
\end{align}
with a constant factor $\cC$ given by
\begin{align}
\cC
&=\Big(\sum_{i}A_i\alpha_i\Big)\frac{(p+q)(q-1)^2}{(p+1)(3q-1)}- \beta \frac{2(q-1)^{2}p!\mathcal{R}_{FFW}}{(p+1)(p+q+1)(3q-1)}
\nonumber
\\
\label{cCdef}
&\quad
+ \gamma \frac{2(q-1)\mathcal{R}_{W^2}}{(p+1)^{2}(p+q+1)(3q-1)}\,.
\end{align}
We will find that the sign of $\cC$ is crucial in the following analysis.

\paragraph{An explicit form for $p=0$.}

To compare our results with the black hole analysis in the literature (see, e.g., Refs.~\cite{Kats:2006xp,Cheung:2018cwt,Hamada:2018dde}), it is convenient to explicitly write down $\cC$ for $p=0$:
\begin{align}
\cC&=(2\alpha_1+\alpha_2)\frac{2q(q-1)^2}{3q-1}
+\beta\frac{2(q-1)^2(q^2-4q+1)}{(q+1)(3q-1)}
+\gamma\frac{4(q-1)(q^3-3q^2+4q-1)}{(q+1)(3q-1)}
\nonumber
\\
&=(2\alpha_1+\alpha_2)\frac{2(d-2)(d-3)^2}{3d-7}
+\beta\frac{2(d-3)^2(d^2-8q+13)}{(d-1)(3d-7)}
\nonumber
\\
\label{C_p0}
&\quad
+\gamma\frac{4(d-3)(d^3-9d^2+28d-29)}{(d-1)(3d-7)}\,,
\end{align}
where we used $q=d-2$ in the second equality. This corresponds to $\mathcal{F}(a_i)$ in Eq. (S26) of Ref.~\cite{Hamada:2018dde} after identification $\alpha_{1,2}\to a_{1,2}$, $\beta\to a_3$ and $\gamma \to a_4$, and setting $e=1$.

\subsection{Extremal condition}

Next we compute higher derivative corrections to the extremal condition.

\paragraph{General consideration.}

In the WGC context, we are interested in the tension-to-charge ration of the zero temperature black branes, so that the canonical ensemble is useful. Let us define the tension-to-charge ratio in the canonical ensemble in the rescaled language as
\begin{align}
\mu(\cT,\cQ)=\frac{\cM(\cT,\cQ)}{\cQ}
=\frac{\cH(\cT,\cQ)+\cT\cS(\cT,\cQ)}{\cQ}\,,
\end{align}
where $\cH(\cT,\cQ)$ is the rescaled Helmholtz free energy in particular. A standard thermodynamic analysis shows that its derivative in $\cQ$ is of the form,
\begin{align}
\nonumber
\frac{\partial \mu(\cT,\cQ)}{\partial \cQ}
&=\frac{1}{\cQ^2}
\left[
\cQ\cPhi(\cT,\cQ)+
\cT\cQ\frac{\partial\cS(\cT,\cQ)}{\partial \cQ}
-\Big(\cH(\cT,\cQ)+\cT\cS(\cT,\cQ)\Big)
\right]
\\
\label{general_mu_Q}
&=\frac{1}{\cQ^2}
\left[-\cG(\cT,\cPhi(\cT,\cQ))+\cT\left(\cQ\frac{\partial\cS(\cT,\cQ)}{\partial \cQ}-\cS(\cT,\cQ)\right)\right]\,,
\end{align}
where $\cG(\cT,\cPhi)$ is the Gibbs free energy in the grand canonical ensemble and $\cPhi(\cT,\cQ)$ is the electric potential in the canonical ensemble. We then define the tension-to-charge ratio of the extremal back brane by
\begin{align}
\mu_{\rm ext}(\cQ)=\mu(0,\cQ)\,.
\end{align}
If the entropy $\cS(\cT,\cQ)$ and its derivative in the electric charge $\cQ$ are finite in the zero temperature limit $\cT\to0$ (it is true in our setup including higher derivative corrections), the second term in Eq.~\eqref{general_mu_Q} vanishes in the extremal limit and therefore
\begin{align}
\label{master_eq}
\frac{d\mu_{\rm ext}(\cQ)}{dQ}=-\frac{\cG(0,\cPhi(0,\cQ))}{\cQ^2}\,.
\end{align}
Our task is now to evaluate the Gibbs free energy in the canonical ensemble and take the zero temperature limit.

\paragraph{Einstein gravity.}

Before applying this consideration to our higher derivative analysis, it is instructive to summarize its implication in the Einstein gravity case. First, let us introduce a dictionary between the grand canonical and canonical ensembles. Recall that the rescaled charge in the grand canonical ensemble is given in Eq.~\eqref{cQ0} as
\begin{align}
\cQ_0(\cT,\cPhi)&=\cPhi\frac{(1-\cPhi^2)^{\frac{q-1}{p+1}}}{\cT^{q-1}}\,.
\end{align}
This gives the rescaled electric potential $\cPhi$ in the canonical ensemble as
\begin{align}
\cPhi_0(\cT,\cQ)=1-\frac{1}{2}\cT^{p+1}\cQ^{\frac{p+1}{q-1}}+\cdots\,,
\end{align}
where the dots stand for higher order terms in $\cT$ that are negligible in the low temperature regime. Also, without loss of generality we assumed that the charge is positive $\cQ>0$. Then, in the extremal limit $\cT\to0$, the Gibbs free energy in the canonical ensemble reads 
\begin{align}
\cG_0(\cT,\cPhi(\cT,\cQ))=\frac{p+1}{2(p+q)}\frac{(1-\cPhi(\cT,\cQ)^2)^{\frac{q-1}{p+1}}}{\cT^{q-1}}(1-\cPhi^2)
\to0\,.
\end{align}
We then conclude that the extremal curve is linear in the Einstein gravity:
\begin{align}
\frac{d\mu_{\rm ext,0}(\cQ)}{d\cQ}=-\frac{\cG_0(0,\cPhi(0,\cQ))}{\cQ^2}=0\,,
\end{align}
which of course agrees with the extremal condition $\mu_{\rm ext,0}(\cQ)=1$ in the Einstein gravity.

\paragraph{Higher derivative analysis.}

Now we move on to our higher derivative analysis. Again, we first introduce a dictionary between the two ensembles. In terms of the higher derivative correction $\Delta\cG$ to the Gibbs free energy evaluated in the previous subsection, we have
\begin{align}
\label{DeltaQ}
\cQ(\cT,\cPhi)=\cQ_0(\cT,\cPhi)
+\Delta \cQ(\cT,\cPhi)
\quad
{\rm with}
\quad
\Delta \cQ(\cT,\cPhi)=-\frac{\partial \Delta\cG(\cT,\cPhi)}{\partial \cPhi}\,.
\end{align}
Then, the electric potential in the canonical ensemble reads
\begin{align}
\cPhi(\cT,\cQ)=\cPhi_0(\cT,\cQ)+\Delta\cPhi(\cT,\cQ)
\end{align}
with the correction term
\begin{align}
\label{Delta_Phi}
\Delta\cPhi(\cT,\cQ)=-\left.\frac{\Delta \cQ(\cT,\cPhi)}{\frac{\partial \cQ_0(\cT,\cPhi)}{\partial\cPhi}}\right|_{\cPhi=\cPhi_0(\cT,\cQ)}
\end{align}
at the leading order. In the extremal limit, higher derivative corrections to the Gibbs free energy in the canonical ensemble are
\begin{align}
\cG(0,\cPhi(0,\cQ))
&\simeq
\cG_0(0,\cPhi_0(0,\cQ))+
\left.\frac{\partial \cG_0(0,\cPhi)}{\partial \cPhi}\right|_{\cPhi=\cPhi_0(0,\cQ)}\Delta \cPhi(0,\cQ)
+\Delta \cG(0,\cPhi_0(0,\cQ))
\nonumber
\\
\label{G_almost}
&=-\cQ\Delta \cPhi(0,\cQ)+\Delta \cG(0,\cPhi_0(0,\cQ))
\,,
\end{align}
where at the second equality we used
\begin{align}
\cG_0(0,\cPhi_0(0,\cQ))=0
\,,
\quad
\left.\frac{\partial \cG_0(0,\cPhi)}{\partial \cPhi}\right|_{\cPhi=\cPhi_0(0,\cQ)}=
-\cQ_0(0,\cPhi_0(0,\cQ))=-\cQ\,.
\end{align}
Using the explicit form~\eqref{DeltaG} of $\Delta\cG(\cT,\cPhi)$ together with Eq.~\eqref{DeltaQ} and Eq.~\eqref{Delta_Phi}, we find
\begin{align}
\cG(0,\cPhi(0,\cQ))
&\simeq  -\frac{2\cC}{q-1}\cQ^{1-\frac{2}{q-1}}\,.
\end{align}
Using the relation~\eqref{master_eq}, we conclude that the correction to the tension-to-charge is
\begin{align}
\label{ext_correction}
\mu_{\rm ext}(\cQ)=1+\Delta \mu_{\rm ext}(\cQ)
\quad
{\rm with}
\quad
\Delta \mu_{\rm ext}(\cQ)\simeq-\cC \cQ^{-\frac{2}{q-1}}\,,
\end{align}
where the integration constant is determined by the extremal condition in the Einstein gravity: $\mu_{\rm ext}(\cQ)\to1$ ($\cQ\to\infty$). The WGC implies that higher derivative corrections decrease the tension-to-charge ratio of extremal black branes, i.e., $\cC>0$.

\medskip
Note that Eq.~\eqref{ext_correction} for $p=0$ reproduces the black hole result in the literature. More explicitly, the charge-to-mass ratio $z_{\rm ext}=\mu_{\rm ext}^{-1}$ of extremal black holes reads
\begin{align}
z_{\rm ext}=1-\Delta\mu_{\rm ext}=1+\cC \cQ^{-\frac{2}{q-1}}
=1+\cC \left(\frac{(d-2)(d-3)V_{d-2}^2}{e^2\kappa^2Q^2}\right)^{\frac{1}{d-3}}\,,
\end{align}
which agrees with Eq. (S24) of Ref.~\cite{Hamada:2018dde} (see below Eq.\eqref{C_p0} for the dictionary between our notation and that of Ref.~\cite{Hamada:2018dde}).

\subsection{Entropy of $\cQ=\cM$ black branes }

It is known that higher derivative corrections to entropy of $\cQ=\cM$ black holes are positive, if and only if corrections to the mass-to-charge ratio of extremal black holes are negative as implied by the WGC (see, e.g., Refs.~\cite{Cheung:2018cwt,Hamada:2018dde,Loges:2019jzs,Goon:2019faz}). In this subsection we explicitly show that the same story holds for black $p$-branes in general spacetime dimensions.

\paragraph{Temperature.}

To discuss higher derivative corrections to thermodynamics of $\cQ=\cM$ black branes, we first identify the temperature $\cT_*(\cQ)$ of the $\cQ=\cM$ black branes as a function of the charge $\cQ$. For this purpose, let us first write down the tension $\cM$ in the canonical ensemble as
\begin{align}
\label{Mmicro}
\cM(\cT,\cQ)
=\cM_0(\cT,\cQ)+\Delta \cM(\cT,\cQ)\,.
\end{align}
Here the first term is the tension in the Einstein gravity, whose explicit form in the low-temperature regime is given by
\begin{align}
\label{M0micro}
\cM_0(\cT,\cQ)=\cQ\left[1+\frac{p(q-1)}{2(p+q)}\epsilon+\frac{2p+1}{8}\epsilon^2+\mathcal{O}(\epsilon^3)\right]
\quad
{\rm with}
\quad
\epsilon=\cQ^{\frac{p+1}{q-1}}\cT^{p+1}
\,.
\end{align}
This shows that $\cQ=\cM$ black branes have zero temperature $\epsilon=0$ in the Einstein gravity. Note that for $p=0$ the second term in the brackets vanishes and so the third term is the next-to-leading order term in the $\epsilon$ expansion.

\medskip
On the other hand, the second term in Eq.~\eqref{Mmicro} is the higher derivative correction to the tension in the canonical ensemble. In the low-temperature limit, it reads
\begin{align}
\Delta \cM(\cT,\cQ)\simeq \Delta \cM(0,\cQ)
=\cQ \Delta \mu_{\rm ext}(\cQ)\,.
\end{align}
This additional contribution to the tension shifts the temperature of $\cQ=\cM$ black branes. More explicitly, at the leading order, $\cM(\cT,\cQ)=\cQ$ implies
\begin{align}
\label{Deltamuext}
-\Delta \mu_{\rm ext}(\cQ)=\left\{\begin{array}{cc}
\displaystyle\frac{1}{8}\epsilon^2 & (p=0)\,,
\\[4mm]
\displaystyle
\frac{p(q-1)}{2(p+q)} \epsilon& (p\geq1)\,.\end{array}\right.
\end{align}
Note that  $\Delta\mu_{\rm ext}(\cQ)< 0$ (which is identical to the WGC bound) is required for the shifted temperature to be real. This is simply because when $\Delta\mu_{\rm ext}(\cQ)>0$, the corrected extremal bound is not satisfied by $\cQ=\cM$, so that $\cQ=\cM$ solutions have no horizon and therefore they are not black branes anymore. Solving Eq.~\eqref{Deltamuext} gives the temperature $\cT_*(\cQ)$ of $\cQ=\cM$ black branes as a function of the charge $\cQ$:
\begin{align}
\label{T*}
\cT_*(\cQ)=\epsilon_*^{\frac{1}{p+1}}\cQ^{-\frac{1}{q-1}}
\quad
{\rm with}
\quad
\epsilon_*(\cQ)=\left\{\begin{array}{cc}
\displaystyle\sqrt{-8\Delta\mu_{\rm ext}(\cQ)}& (p=0)\,,
\\[2mm]
\displaystyle
-\frac{2(p+q)}{p(q-1)} \Delta\mu_{\rm ext}(\cQ)& (p\geq1)\,.\end{array}\right.
\end{align}
Since Eqs.~\eqref{Deltamuext}-\eqref{T*} are qualitatively different for $p\geq 1$ and $p=0$, below we study the two cases separately.

\subsubsection{Black branes with $p\geq1$}

We start from the $p\geq1$ case and evaluate higher derivative corrections to the electric potential, Gibbs free energy, and then entropy, respectively.

\paragraph{Electric potential.}

Let us write the electric potential of the $\cQ=\cM$ black branes as
\begin{align}
\cPhi_*(\cQ)&=\cPhi_0(\cT_*(\cQ),\cQ)+\Delta \cPhi(\cT_*(\cQ),\cQ)\,.
\end{align}
Here $\cPhi_0(\cT,\cQ)$ is the electric potential in the canonical ensemble for the Einstein gravity, which is given by
\begin{align}
\cPhi_0(\cT_*(\cQ),\cQ)=1-\frac{1}{2}\epsilon_*(\cQ)+\mathcal{O}(\epsilon_*^2)\,.
\end{align}
On the other hand, $\Delta \cPhi(\cT,\cQ)$ is the higher derivative correction, whose low-temperature behavior reads
\begin{align}
\Delta \cPhi(\cT_*(\cQ),\cQ)\simeq\frac{q-3}{q-1}\cC\cQ^{-\frac{2}{q-1}}=-\frac{q-3}{q-1}\Delta\mu_{\rm ext}(\cQ)
=\frac{p(q-3)}{2(p+q)}\epsilon_*(\cQ)\,.
\end{align}
We then find
\begin{align}
\label{Phi*1}
\cPhi_*(\cQ)=1-\frac{1}{2}\left(
1-\frac{p(q-3)}{p+q}\right)\epsilon_*(\cQ)+\mathcal{O}(\epsilon_*&^2)\,.
\end{align}

\paragraph{Gibbs free energy.}

Similarly, we decompose the Gibbs free energy of $\cQ=\cM$ black branes as
\begin{align}
\cG_*(\cQ)=\cG_0(\cT_*(\cQ),\cPhi_*(\cQ))+\Delta\cG(\cT_*(\cQ),\cPhi_*(\cQ))\,.
\end{align}
Here the first term is the Einstein gravity result:
\begin{align}
\cG_0(\cT_*(\cQ),\cPhi_*(\cQ))=\cQ\left[
\frac{(1+p)}{2(p+q)}\epsilon_*(\cQ)+\mathcal{O}(\epsilon^2)
\right]\,.
\end{align}
The second term is the higher derivative correction, whose low-temperature behavior is
\begin{align}
\Delta\cG(\cT_*(\cQ),\cPhi_*(\cQ))\simeq
-\cC\cQ^{1-\frac{2}{q-1}}=\cQ \Delta\mu_{\rm ext}(\cQ)
=
-\frac{p(q-1)}{2(p+q)}\cQ\epsilon_*(\cQ) \,.
\end{align}
Therefore, we have
\begin{align}
\label{G*1}
\cG_*(\cQ)=
\cQ
\left[
\frac{1+2p-pq}{2(p+q)}\epsilon_*(\cQ)
+\mathcal{O}(\epsilon_*^2)
\right]
\,.
\end{align}

\paragraph{Entropy.}

Finally, we compute the entropy of $\cQ=\cM$ black branes. For this purpose, it is convenient to note that the thermodynamic relation $\cM=\cG+\cT\cS+\cQ\cPhi$ implies that the entropy of $\cQ=\cM$ black brane is
\begin{align}
\cS_*(\cQ)=\frac{(1-\cPhi_*(\cQ))\cQ-\cG_*(\cQ)}{\cT_*(\cQ)}
\,.
\end{align}
Using Eqs.~\eqref{T*}, \eqref{Phi*1}, and~\eqref{G*1}, we find
\begin{align}
\cS_*(\cQ)&=\frac{2p+q-1}{2(p+q)}\cQ^{\frac{q}{q-1}}\epsilon_*^{\frac{p}{p+1}}
\left[
1+\mathcal{O}(\epsilon_*)
\right]
\,,
\end{align}
which is consistent with the fact that extremal black branes with $p\geq1$ have vanishing entropy $\cS_{*0}(\cQ)=0$. We also find that entropy correction is positive as long as $\Delta\mu_{\rm ext}(\cQ)<0$ and therefore $T_*(\cQ)$ is real and $\epsilon_*(\cQ)>0$. A direct relation between  entropy correction $\Delta\cS_*(\cQ)=\cS_*(\cQ)-\cS_{0*}(\cQ)$ and the correction to the extremal bound $\Delta \mu_{\rm ext(\cQ)}$ reads
\begin{align}
\Delta \cS_*(\cQ)\simeq
\frac{2p+q-1}{2(p+q)}\left(\frac{2(p+q)}{p(q-1)}\right)^{\frac{p}{p+1}}\cQ^{\frac{q}{q-1}}\Big(- \Delta\mu_{\rm ext}(\cQ)\Big)^{\frac{p}{p+1}}
\end{align}
at the leading order.

\subsubsection{Black holes ($p=0$)}

The analysis for $p=0$ can also be performed in the same manner. The only thing to take care here is that $\Delta\mu_{\rm ext}(\cQ)=\mathcal{O}(\epsilon_*^2)$ in contrast to the $p\geq1$ case, where $\Delta\mu_{\rm ext}(\cQ)=\mathcal{O}(\epsilon_*)$. See Eq.~\eqref{T*}. Accordingly, we include the next order of the $\epsilon_*$ expansion in the analysis.

\paragraph{Electric potential.}

First, the electric potential of $\cQ=\cM$ black holes reads
\begin{align}
\cPhi_*(\cQ)&=\cPhi_0(\cT_*(\cQ),\cQ)+\Delta \cPhi(\cT_*(\cQ),\cQ)
\end{align}
with
\begin{align}
\cPhi_0(\cT_*(\cQ),\cQ)&=1-\frac{1}{2}\epsilon_*(\cQ)-\frac{q+1}{8(q-1)}\epsilon_*(\cQ)^2+\mathcal{O}(\epsilon_*^3)\,,
\\
\Delta \cPhi(\cT_*(\cQ),\cQ)&\simeq\frac{q-3}{q-1}\cC\cQ^{-\frac{2}{q-1}}=-\frac{q-3}{q-1}\Delta\mu_{\rm ext}(\cQ)
=\frac{q-3}{8(q-1)}\epsilon_*(\cQ)^2\,.
\end{align}
We then have
\begin{align}
\cPhi_*(\cQ)=1-\frac{1}{2}\epsilon_*(\cQ)-\frac{1}{2(q-1)}\epsilon_*(\cQ)^2+\mathcal{O}(\epsilon_*^3)
\,.
\end{align}

\paragraph{Gibbs free energy.}

Similarly, the Gibbs free energy of $\cQ=\cM$ black holes is
\begin{align}
\cG_*(\cQ)=\cG_0(\cT_*(\cQ),\cPhi_*(\cQ))+\Delta\cG(\cT_*(\cQ),\cPhi_*(\cQ))
\end{align}
with
\begin{align}
\cG_0(\cT_*(\cQ),\cPhi_*(\cQ))&=\cQ\left[
\frac{1}{2q}\epsilon_*(\cQ)-\frac{q-5}{8(q-1)}\epsilon_*(\cQ)^2+\mathcal{O}(\epsilon^3)
\right]\,,
\\
\Delta\cG(\cT_*(\cQ),\cPhi_*(\cQ))&\simeq
-\cC\cQ^{1-\frac{2}{q-1}}=\cQ \Delta\mu_{\rm ext}(\cQ)=-\frac{1}{8}\epsilon_*(\cQ)^2\,.
\end{align}
Therefore, we have
\begin{align}
\cG_*(\cQ)=
\cQ
\left[
\frac{1}{2q}\epsilon_*(\cQ)
-\frac{q-3}{4(q-1)}\epsilon_*(\cQ)^2+\mathcal{O}(\epsilon_*^3)
\right]
\,.
\end{align}

\paragraph{Entropy.}

Finally, the entropy of $\cQ=\cM$ black holes reads
\begin{align}
\nonumber
\cS_*(\cQ)&=\frac{(1-\cPhi_*(\cQ))\cQ-\cG_*(\cQ)}{\cT_*(\cQ)}
\\
\nonumber
&=\frac{\cQ}{\cT_*(\cQ)}
\left[\frac{q-1}{2q}\epsilon_*(\cQ)
+\frac{1}{4}\epsilon_*(\cQ)^2+\mathcal{O}(\epsilon_*^3)
\right]
\\
&=\cQ^{\frac{q}{q-1}}
\left[
\frac{q-1}{2q}+\frac{1}{4}\epsilon_*(\cQ)+\mathcal{O}(\epsilon_*^2)
\right]
\,,
\end{align}
where the first term reproduces the entropy of $\cQ=\cM$ black holes $\displaystyle\cS_{*0}(\cQ)=\frac{q-1}{2q}\cQ^{\frac{q}{q-1}}$. Then, the entropy correction reads
\begin{align}
\Delta\cS_*(\cQ)=\cS_*(\cQ)-\cS_{*0}(\cQ)
\simeq
\frac{1}{4}\epsilon_*(\cQ)\cQ^{\frac{q}{q-1}}\,,
\end{align}
which is positive as long as $\Delta\mu_{\rm ext}(\cQ)<0$ and therefore $T_*(\cQ)$ is real and $\epsilon_*(\cQ)>0$. A more direct relation between $\Delta \cS_*$ and $\Delta \mu_{\rm ext}$ is
\begin{align}
\Delta\cS_*(\cQ)\simeq
\frac{1}{\sqrt2}\sqrt{-\Delta\mu_{\rm ext}(\cQ)}\,\cQ^{\frac{q}{q-1}}\,,
\end{align}
which agrees with the known result in the literature~\cite{Hamada:2018dde}\footnote{
The agreement is after correcting typos in Ref.~\cite{Hamada:2018dde}. First, Eq. (S53) has to be corrected as
\begin{align}
\frac{\Delta r_H^2}{r_H^2}=-\frac{2\Delta g(r_H)}{r_H^2g''_{EM}(r_H)}=\frac{2\mathcal{F}(a_i)}{(D-3)^2m^{\frac{2}{D-3}}}\,.
\end{align}
Accordingly, Eq. (S54) should be modified as
\begin{align}
\frac{\Delta S}{S_{EM}}\simeq\frac{\Delta S_{\rm horizon}}{S_{EM}}
=\frac{\sqrt{2}(D-2)}{(D-3)m^{\frac{1}{D-3}}}\sqrt{\mathcal{F}(a_i)}\,.
\end{align}
}.

\subsection{Implications of positivity bounds}
\label{Sec:positivity}

So far, we have shown that positive sign of $\cC$ given in Eq.~\eqref{cCdef} is favored by the WGC and also the same sign follows from positivity of the entropy correction of $\cM=\cQ$ black branes. In the following, we conclude our higher derivative analysis by demonstrating that the same sign is favored by positivity bounds on scattering amplitudes~\cite{Adams:2006sv}.

\paragraph{Positivity bounds and their limitation.}

Positivity bounds provide a necessary condition for an IR scattering amplitude to have a consistent UV completion that respects unitarity, analyticity and locality~\cite{Adams:2006sv}. They are formulated in terms of the IR expansion of an $s$-$u$ symmetric scattering amplitude $M(s,t)$. Here for notational simplicity we assume that external particles are massless. If the amplitude in the forward limit enjoys the IR expansion of the form,
\begin{align}
\label{IR_expansion}
\mathcal{M}(s,t=0)=\sum_{n=0}^\infty a_{2n}s^{2n}\,,
\end{align}
the UV assumptions mentioned earlier imply that the coefficients $a_{2n}$ of $s^{2n}$ ($2n=2,4,\ldots$) are all positive, which are called the positivity bounds.

\medskip
However, the story changes in the presence of gravity because the IR expansion~\eqref{IR_expansion} does not work anymore. For example, tree-level graviton exchange gives a $t$-channel pole  $\displaystyle\sim \kappa^2\frac{s^2}{t}$, which dominates over the $s^2$ term in the forward limit. As a consequence, positivity of the $s^2$ coefficient does not follow straightforwardly. The best we can do in the present technology is to show that positivity of the $s^2$ coefficient holds approximately as long as gravitational effects are subdominant. See, e.g., Refs.~\cite{Hamada:2018dde,Bellazzini:2019xts,Alberte:2020jsk,Tokuda:2020mlf,Herrero-Valea:2020wxz,Arkani-Hamed:2020blm,Caron-Huot:2021rmr} for more details.

\paragraph{Positivity vs $\cC>0$.}

Having said this caveat, we discuss implications of the approximate positivity of the $s^2$ coefficient for our higher derivative analysis. Since the approximate positivity works when gravity is subdominant, here we focus on the situation where contributions of $FFW$ and $W^2$ are subdominant compared to the $F^4$ term, i.e.,
\begin{align}
\cC\simeq\Big(\sum_{i}A_i\alpha_i\Big)\frac{(p+q)(q-1)^2}{(p+1)(3q-1)}\,.
\end{align}
Below we show that the approximate positivity implies $\sum_{i}A_i\alpha_i
\gtrsim 0$ and therefore $\cC\gtrsim0$.\footnote{
We use $\gtrsim$ rather than $>$ to emphasize that known positivity bounds on the $s^2$ coefficient are only approximate in the presence of gravity.}

\medskip
For this purpose, we consider four-point scattering of the $(p+1)$-form $A$ and decompose the spacetime coordinates into $3$ dimensions $x^a$ ($a=0,1,2$) on the scattering plane and $(d-3)$ dimensions $x^{I}$ $(I=3,\ldots,d-1)$ orthogonal to it. Also let us suppose that all the external particles are polarized on the same $(p+1)$-dimensional plane, say on the plane $(x^3,x^4,\ldots, x^{p+3})$. By analogy with the Kaluza-Klein decomposition, we identify those external modes with a scalar field $\phi(x^a)$ on the $3$-dimensional scattering plane:
\begin{align}
\phi(x^a)=A_{3\,4\,\ldots \,p+3}(x^a)\,.
\end{align}
Then, the scattering amplitude can be computed by rewriting the effective action~\eqref{Lorentzian_action} in terms of $\phi$ (and other modes relevant to our scattering problem, if any). If we ignore gravity by our working assumption, the only interaction relevant to four-point scattering of $\phi$ is  $(\partial_a\phi\partial^a\phi)^2$ originating from the $F^4$ operator.\footnote{
As we mentioned earlier, when $p+1=4n$ ($n=1,2,\ldots$), the effective action contains three-derivative operators of the form $F^3$, but they do not play any role in four-point scattering of $\phi$ at the tree-level.}
Noticing that $F_{a\,3\,4\,\ldots\,p+3}=\partial_a\phi$, the effective Lagrangian relevant to our scattering problem is
\begin{align}
\mathcal{L}=-\frac{1}{2}(\partial_a\phi)^2+\frac{\kappa^2}{e^4}\Big(\sum_{i}A_i\alpha_i\Big)(\partial_a\phi\partial^a\phi)^2\,.
\end{align}
We emphasize that the same combinatorial factor $A_i$ defined in Eq.~\eqref{F^4} appears in the effective Lagrangian. Then, the approximate positivity bound on the $s^2$ coefficient reads
\begin{align}
\sum_{i}A_i\alpha_i\gtrsim0\,.
\end{align}
To summarize, we have shown that when gravity is subdominant, the approximate positivity on the $s^2$ coefficient implies $\sum_{i}A_i\alpha_i\gtrsim0$ and therefore $\cC\gtrsim0$ as expected by the WGC and positivity of the entropy correction of $\cQ=\cM$ black branes.

\section{Discussion}

In this paper we studied higher derivative corrections to black brane thermodynamics and their implications for the WGC. We found that higher derivative corrections decrease the tension-to-charge ratio of extremal black branes as expected by the WGC, if and only iff the constant factor $\cC$ defined in Eq.~\eqref{cCdef} is positive. In particular, we demonstrated that the positive sign is favored by the approximate positivity bounds on scattering amplitudes. We also found that for the same sign of $\cC$, the entropy correction of $\cQ=\cM$ black branes is positive. This extends earlier works in the Einstein-Maxwell theory to $p$-form gauge fields in general spacetime dimensions. 

\medskip
There are various future directions along the line of our present work. First, it is natural to extend our analysis to $p$-form gauge fields coupled to dilaton and axion. There will be no obstruction to the thermodynamic analysis besides technical complication. For example, it would be interesting to study the role of duality constraints in this context. Another interesting direction is to go beyond the leading order correction. In the context of the WGC for $1$-form gauge fields, it has been studied how to interpolate the extremal curve of large black holes and the charged state spectrum below the Planck scale, e.g., based on modular invariance of the world-sheet and/or the dual CFT~\cite{Heidenreich:2016aqi,Montero:2016tif,Aalsma:2019ryi}. It would be interesting to generalize the discussion to $p$-form gauge fields. Moreover, it would be great if we can reinterpret the earlier implication of modular invariance from the thermodynamic perspective. Our question is what the thermodynamics would tell us about microscopic back holes/branes and the subPlanckian spectrum.

%
%Our work leaves a number of ways to future work. First, it is natural to extend our discussion to Einstein-$p$-form-dilaton theory. A dilaton does not change our discussion in Sec.~\ref{Sec:thermo} and Sec.~\ref{Sec:corrections}, and the correction to charge-to-tension ratio of black brane can be calculated systematically. With a dilaton the low-energy effective field theory includes more degrees of freedom, and only the scattering positivity bound can not ensure the weak gravity bound as shown in \cite{Loges:2019jzs} for 4d and 1-form case. We need to found the other assumptions on the UV theory for genaral dimensions. Second, how our discussion of mild form of WGC is generalized to of the subLattic/Tower WGC? We assumed that the tension of black branes $M$ are more larger than the Planck mass $M_{\text{Pl}}$ and we can ignore more higher derivative terms like $(F^{2})^{3}$. These assumptions allow us to calculate easily corrections to the charge-to-tension ratio. However, in the case of $M\sim M_{Pl}$, we must consider more higher derivative terms, and these terms change our discussion in Sec~\ref{Sec:corrections}. Our question is how these terms effect thermodynamics of black branes.

\medskip
\begin{acknowledgments}

We would like to thank Yoshihiko Abe, Gregory J. Loges and Gary Shiu for useful discussion.
T.N. is supported in part by JSPS KAKENHI Grant No.~20H01902 and No.~22H01220, and MEXT KAKENHI Grant No.~21H00075, No.~21H05184 and No.~21H05462.

\end{acknowledgments}

\appendix

\section{Curvature tensors for black brane solutions}
\label{App:curvature}

\paragraph{Riemann tensor.}

For the black brane solution~\eqref{BB1} in the Einstein gravity, nontrivial components of the Riemann tensor are given in terms of the functions $f_\pm(r)$ as
\begin{align*}
		R^{t}_{\ rtr} &= -\frac{1}{2}\left(\frac{f_{+}''}{f_{+}}-\frac{p-2}{p+1}\frac{f_{+}'}{f_{+}}\frac{f_{-}'}{f_{-}}-\frac{p-1}{p+1}\frac{f_{-}''}{f_{-}}+\frac{p-1}{p+1}\frac{p}{p+1}\frac{f_{-}'^{2}}{f_{-}^{2}}\right)\,,	\\
	R^{t}_{\ \theta_{i}t\theta_{i}} &= -\frac{r}{2}\left(f_{+}'f_{-}-\frac{p-1}{p+1}f_{+}f_{-}'\right)\prod_{k=1}^{i-1}\sin^{2}\theta_{k}\,,	\\
	R^{t}_{\ y^{i}ty^{i}} &= -\frac{1}{2(p+1)}\left(f_{+}'f_{-}-\frac{p-1}{p+1}f_{+}f_{-}'\right)f_{-}'f_{-}^{-\frac{p-1}{p+1}}\,,	\\
	R^{r}_{\ \theta_{i}r\theta_{i}} &= -\frac{r}{2}\left(f_{+}'f_{-}+f_{+}f_{-}'\right)\prod_{k=1}^{i-1}\sin^{2}\theta_{k}\,,	\\
	R^{r}_{\ y^{i}ry^{i}} &= -\frac{1}{2(p+1)}\left(f_{+}'f_{-}' + 2f_{+}f_{-}''-\frac{p-1}{p+1}f_{+}\frac{f_{-}'^{2}}{f_{-}}\right)f_{-}^{\frac{2}{p+1}}\,,\\
	R^{\theta_{i}}_{\ \theta_{j}\theta_{i}\theta_{j}} &= (1-f_{+}f_{-})\prod_{k=1}^{j-1}\sin^{2}\theta_{k}\qquad(i>j)\,,	\\
	R^{\theta_{i}}_{\ y^{j}\theta_{i}y^{j}} &= -\frac{1}{p+1}\frac{1}{r}f_{+}f_{-}'f_{-}^{\frac{2}{p+1}}\,,\\
	R^{y^{i}}_{\ y^{j}y^{i}y^{j}} &= -\frac{1}{(p+1)^{2}}f_{+}f_{-}'^{2}f_{-}^{-\frac{p-1}{p+1}}\qquad(i\neq j)\,.
\end{align*}

\paragraph{Ricci curvatures.}

Nonzero components of the Ricci tensor and the Ricci scalar are
\begin{align*}
	R_{tt} &= \frac{1}{2}f_{+}f_{-}^{-\frac{p-1}{p+1}}\left[f_{+}''f_{-}+\frac{2}{p+1}f_{+}'f_{-}'-\frac{p-1}{p+1}f_{+}f_{-}''+\frac{q}{r}\left(f_{+}'f_{-}-\frac{p-1}{p+1}f_{+}f_{-}'\right)\right]\,,\\
	R_{rr} &= -\frac{1}{2}f_{+}^{-1}f_{-}^{-1}\left[f_{+}''f_{-}+\frac{2}{p+1}f_{+}'f_{-}'+f_{+}f_{-}''+\frac{q}{r}\left(f_{+}'f_{-}+f_{+}f_{-}'\right)\right]\,,\\
	R_{\theta_{i}\theta_{i}} &= -\left[r\left(f_{+}'f_{-}+f_{+}f_{-}'\right)-(q-1)\left(1-f_{+}f_{-}\right)\right]\prod_{k=1}^{i-1}\sin^{2}\theta_{k}\,,\\
	R_{y^{i}y^{i}} &= -\frac{1}{p+1}f_{-}^{\frac{2}{p+1}}\left[f_{+}'f_{-}'+f_{+}f_{-}''+\frac{q}{r}f_{+}f_{-}'\right]\,,	\\
	R &= \frac{q(q-1)}{r^{2}}\left(1-f_{+}f_{-}\right)-2\frac{q}{r}\left(f_{+}'f_{-}+f_{+}f_{-}'\right)	
	 -\left(f_{+}''f_{-}+\frac{p+2}{p+1}f_{+}'f_{-}'+f_{+}f_{-}''\right)\,.
\end{align*}

\paragraph{Weyl tensor.}

The Weyl tensor is defined in spacetime $d$ dimensions as
\begin{align}
	W_{\mu\nu\rho\sigma} &= R_{\mu\nu\rho\sigma} + \frac{1}{d-2}\left(R_{\mu\sigma}g_{\nu\rho}-R_{\mu\rho}g_{\nu\sigma}+R_{\nu\rho}g_{\mu\sigma}-R_{\nu\sigma}g_{\mu\rho}\right)	\notag\\*
	&\qquad +\frac{1}{(d-2)(d-1)}R(g_{\mu\rho}g_{\nu\sigma}-g_{\mu\sigma}g_{\nu\rho})\,.	\label{Weyl}
\end{align} 
Its square can be reformulated in terms of the Riemann tensor as 
\begin{equation}
	W^{2} \equiv W_{\mu\nu\rho\sigma}W^{\mu\nu\rho\sigma} = R_{\mu\nu\rho\sigma}^{2} - \frac{4}{d-2}R_{\mu\nu}^{2} + \frac{2}{(d-1)(d-2)}R^{2}\,,
\end{equation}
which can be evaluated for the solution~\eqref{BB1} in a straightforward manner using the formulae provided in this appendix.

\section{An explicit form of $\Delta\cG_{W^2}$}
\label{app:GWW}

An explicit form of $\Delta \cG_{W^2}$ including subleading terms omitted in Eq.~\eqref{delatWW} is
\begin{align}
\Delta\cG_{W^2}&=
-\gamma\, \widetilde{\cC}_{W^2}\cT^{3-q}\left(1-\cPhi^{2}\right)^{\frac{q-3}{p+1}}
\end{align}
with $\widetilde{\cC}_{W^2}$ defined by
\begin{align}
\widetilde{\cC}_{W^2}
&=-\frac{p(p-1)q(q-1)^3}{(p+1)^2(p+q)}
\cPhi^{-2}
+\mathcal{P}_{0}+\mathcal{P}_{2}\cPhi^{2}
+\mathcal{P}_{4}\cPhi^{4}
\nonumber
\\
\label{Ctilde}
&\quad
+\frac{p(p-1)q(q-1)^3}{(p+1)^2(p+q)}\frac{(1-\cPhi^2)^2}{\cPhi^2}{}_{2}F_{1}\left(1,\frac{2}{q-1};1+\frac{2}{q-1};\cPhi^2\right)\,.
\end{align}
Here $\mathcal{P}_{0,2,4}$ are $(p,q)$-dependent factors given by
\begin{align*}
\mathcal{P}_{0}&=\frac{q^2(q-1)}{(p+1)^2(p+q)(q+1)}
\left[
p^3(q+1)
+p^2(2q^2-q+5)
+p(-2q^2+7q+1)
+(q+1)
\right]\,,
\\
\mathcal{P}_{2}&=-\frac{(q-1)^2}{(p+1)^2(q+1)}
\left[
p^2(q+1)^2
-p(q^2-8q-1)
+2(q+1)
\right]\,,
\\
\mathcal{P}_{4}&=\frac{(q-1)^2}{(p+1)^2(p+q+1)(q+1)(3q-1)}
\left[
p^3(4q-1)(q+1)
+2p^2(2q^3+7q^2-2q+1)
\right.\\
&\qquad\qquad\qquad\qquad\qquad\qquad\qquad
\left.
+5p(q+1)(3q^2-3q+1)
+(q-1)(q+1)(7q-2)
\right]\,.
\end{align*}
Notice in particular that the second line of Eq.~\eqref{Ctilde} vanishes in the limit $\cPhi^2\to1$.

\bibliography{braneWGC}{}
\bibliographystyle{utphys}

\end{document}